\documentclass[twocolumn,showpacs,preprintnumbers,amsmath,amssymb]{revtex4-1}
\usepackage{epsfig}
\usepackage{graphicx}
\usepackage{epsfig}
\usepackage{color}
\usepackage{rotating}
\usepackage{amssymb}
\usepackage{amsmath}
\usepackage{graphicx}

\def\beq{\begin{eqnarray}}
\def\eeq{\end{eqnarray}}

\newcommand{\be}{\begin{equation}}
\newcommand{\ee}{\end{equation}}
\newcommand{\bea}{\begin{eqnarray}}
\newcommand{\eea}{\end{eqnarray}}

\begin{document}

\title{On work and heat in time-dependent strong coupling}

\author{Erik Aurell}
\email{eaurell@kth.se}
\affiliation{
KTH -- Royal Institute of Technology,  AlbaNova University Center, SE-106 91~Stockholm, Sweden\\
Depts. Computer Science and Applied Physics, Aalto University, Espoo, Finland
}

\begin{abstract}
This paper revisits the classical problem of representing 
a thermal bath interacting with a system as a large collection of harmonic oscillators initially in thermal
equilibrium. As is well known the system then obeys an equation, which in the bulk and in the
suitable limit tends to the Kramers-Langevin equation of physical kinetics.
I consider time-dependent system-bath coupling and show that
this leads to an additional harmonic force acting on the system.
When the coupling is switched on and switched off rapidly 
the force has delta-function support at the initial and final time,
an effect that has been observed previously. 
I further show that the work and heat functionals as recently defined in
stochastic thermodynamics at strong coupling contain additional terms 
depending on the time derivative of the system-bath coupling.
I discuss these terms and
show that while they can be very large 
if the system-bath coupling changes quickly
they only give a finite contribution to the work that enter in Jarzynski's equality. 
I also discuss that these corrections to standard work and heat
functionals provide an explanation for
non-standard terms in the  
change of the von Neumann entropy of a quantum bath
interacting with a quantum system found in an earlier contribution (Aurell \& Eichhorn, 2015).
\end{abstract}


\maketitle

\section{Introduction}
\label{sec:Introduction}
Consider a collection of $N$ harmonic oscillators governed by the Hamiltonian
\begin{equation}
H_B = \sum_b \frac{p_b^2}{2m_b}+\frac{1}{2}m_b\omega_b^2 q_b^2.
\label{eq:H_B}
\end{equation}
The oscillators start in thermal equilibrium at some inverse temperature $\beta$. We will call this the bath.
Consider also another externally driven system described by a Hamiltonian
\begin{equation}
H_S = \frac{P^2}{2M}+V(Q,t)
\label{eq:H_S}
\end{equation}
and consider the case when the system and the bath interact linearly during some finite time interval.
The interaction Hamiltonian is
\begin{equation}
H_I(t) = - \sum_b C_b(t)q_b Q 
\label{eq:H_I}
\end{equation}
where the $C_b(t)$ are functions of time, and to this we add the counter-term Hamiltonian
\begin{equation}
H_C(t) = \sum_b \frac{C_b^2(t)Q^2}{2m_b\omega_b^2}.
\label{eq:H_C}
\end{equation}
The total Hamiltonian of the bath and the system is
\begin{equation}
H_{TOT} = H_S + H_B + H_I + H_C.
\label{eq:H_TOT}
\end{equation}

We will be concerned with the setting
that $C_b(t)$ rise rapidly from zero to some steady values $C_b$ after an initial time $t_i$
and decrease rapidly back to zero before some final time $t_f$. 
The system and the bath are therefore disconnected before and at time $t_i$ as well as at and after time $t_f$.
It is natural to postulate that the heat exchanged between the system and the bath
during the process is $\Delta H_B$, the change of the energy of the bath
in the time interval $[ t_i:t_f ]$.



The paper has two goals.
The first is to compare $\Delta H_B$ to heat, work and internal energy in the recently introduced Stochastic Thermodynamics at strong coupling~\cite{Seifert2016,Jarzynski2017}.
The new effect that will appear is a work 
which arises from the change of the system-bath coupling, and which is entirely dissipated into heat.
When the change of the system-bath coupling is fast this work is large, and can
pathwise and on average be (much) larger than the standard Jarzynski work.
Due to a cancellation effect most of it however does not contribute to the average in Jarzynski's equality. 
The second goal is to explain results on the classical limit of the corresponding quantum problem
which were found in an earlier contribution~\cite{AurellEichhorn}.
It will be shown that the main extra term found there was an artefact from assuming
that the total density matrix of the system and the bath being initially factorized,
while the system and the bath are nevertheless interacting from the beginning of the process.
If instead the system-bath coupling goes to zero smoothly at the beginning and the end of
the process the classical limit of heat is recovered fully, and quantum-classical
correspondence holds on the level of expectation values. 

The paper is organized as follows. In Sections~\ref{sec:Zwanzig} and~\ref{sec:Ohmic} I describe
how interaction with a bath as defined here leads to a Kramers-Langevin dynamics for the system,
with the additional terms that appear when the system-bath coupling depends on time.
In Section~\ref{sec:StochThermo} I 
consider work, internal energy and heat as recently defined in Stochastic Thermodynamics at strong coupling
and relate them to $\Delta H_B$, and 
in Section~\ref{sec:S} I show how these functionals of the system history
are modified when the system-bath coupling depend on time.
In Section~\ref{sec:JE} I discuss the extra work referred to above,
and in Section~\ref{sec:Quantum} I 
compare the expectation value of $\Delta H_B$ to the earlier computed classical limit of the change of the von Neumann entropy of the bath. In Section~\ref{sec:Discussion} I sum up the results.
In Appendix~\ref{a:conditioning} I discuss for completeness an 
additional term in the classical limit of the change
of the von Neumann entropy that was also found in~\cite{AurellEichhorn}, but which has 
a different origin than the main focus of this paper.
 
\section{Recall of the Zwanzig theory}
\label{sec:Zwanzig}
All the calculations in this section are straightforward, and mostly re-statements of known results~\cite{Zwanzig}.
The equations of motion of the bath and the system that follow from
(\ref{eq:H_B}), (\ref{eq:H_S}) (\ref{eq:H_I}) and (\ref{eq:H_C}) are 
\begin{eqnarray}
\label{eq:pb-dot}
\dot{p_b} &=& -m_b\omega_b^2 q_b + C_b(t)Q\\
\label{eq:qb-dot}
\dot{q_b} &=& \frac{p_b}{m_b}\\
\label{eq:P-dot}
\dot{P}   &=& -V'(Q,t) + \sum_b C_b(t)\left(q_b- \frac{C_b(t)}{m_b\omega_b^2}Q\right)\\
\label{eq:Q-dot}
\dot{Q}   &=& \frac{P}{M}
\end{eqnarray} 
The initial position and momentum of the system $(Q^i,P^i)$ are assumed known. All 
randomness in this problem is hence due to lack of knowledge of the initial positions and momenta of the bath.
These are $2N$ real numbers while the time history of the system over the finite time interval,
$\{Q,P\}_{t_i}^{t_f}$, needs infinitely many variables to be completely specified. 
One may therefore assume that a complete knowledge of the system history uniquely specifies the 
initial positions and momenta of the bath; to the equilibrium probability $P^{eq}$ in the bath 
then corresponds a probability distribution $P\left[\{Q,P\}_{t_i}^{t_f}\right]$ on system histories. 
This will be assumed throughout the following.

The equations of motion for the bath oscillators can be solved as
\begin{widetext}
\begin{eqnarray}
q_b(t) &=& \left(q_b(t_i)-\frac{C_b(t_i)Q(t_i)}{m_b\omega_b^2}\right)\cos\omega_b(t-t_i)  + \frac{C_b(t)Q(t)}{m_b\omega_b^2} -\nonumber \\
       && \frac{1}{\omega_b^2 m_b} \int_{t_i}^t [\cos(\omega_b(t-s)) \left(\dot{C}_b(s) Q(s)+C_b(s) \dot{Q}(s)\right)] ds +\, p_b(t_i)\frac{1}{\omega_b m_b}\sin\omega_b(t-t_i)
\label{eq:Ralf-solution} 
\end{eqnarray}
\end{widetext}
By the assumption that the system and the bath are initially disconnected $C_b(t_i)Q(t_i)$ vanishes.
Inserting $q_b(t)$ in the equations of motion for the system give
\begin{eqnarray}
\dot{Q} &=& \frac{P}{M}\qquad \dot{P} = -V^{\prime}(Q,t) + \xi(t) \nonumber \\  
&&\qquad - \int_{t_i}^t \gamma(t,s)\dot{Q}(s)+\chi(t,s)Q(s) ds 
\label{eq:FP-1}
\end{eqnarray}
using the ancillary quantities
\begin{eqnarray}
\label{eq:xi}
\xi(t)    &=& \sum_b C_b(t) \big[ q_b(t_i)\cos\omega_b(t-t_i) + \nonumber \\
        &&\quad p_b(t_i)\frac{1}{\omega_b m_b}\sin\omega_b(t-t_i) \big] \\ 
\label{eq:gamma}
\gamma(t,s)&=& \sum_b \frac{C_b(t)C_b(s)}{m_b\omega_b^2} \cos\omega_b(t-s)  \\
\label{eq:chi}
\chi(t,s)&=& \sum_b \frac{C_b(t)\dot{C}_b(s)}{m_b\omega_b^2}\cos\omega_b(t-s) 
\end{eqnarray}
Equations (\ref{eq:FP-1}) and (\ref{eq:xi}-\ref{eq:chi}) are the solutions
to the general problem of a system interacting linearly with a bath of
harmonic oscillators.

\section{The time-dependent Ohmic bath and its forces}
\label{sec:Ohmic}
In this section we introduce additional assumptions so that
(\ref{eq:FP-1}) is classic Kramers-Langevin equation
when the system-bath coupling is constant.  
The starting point is to follow Caldeira and Leggett~\cite{CaldeiraLeggett83a} 
and assume an Ohmic bath \textit{i.e.} that the spectrum of the bath oscillators is continuous up to an upper cut-off $\Omega$
and increases quadratically with frequency. The number of oscillators with frequencies in the
interval $[\omega,\omega+d\omega]$ is $f(\omega)d\omega$ and 
\begin{equation}
 \begin{array}{lcll}
     f(\omega) &=& \frac{2}{\pi}\omega_c^{-3}\omega^2 \qquad &\omega<\Omega \\
     f(\omega) &=& 0                            & \omega>\Omega 
 \end{array}    
\end{equation}
where $\omega_c$ is some characteristic frequency less than $\Omega$.
The total number of oscillators is hence $\frac{2}{3\pi}\left(\frac{\Omega}{\omega_c}\right)^3$.
We also assume a function $\eta(s)$ with dimension (mass/time) such that 
\begin{equation}
\label{eq:C-omega}
C_{\omega}(s)=\sqrt{\omega_c^{3}m_{\omega}\eta(s)}
\end{equation}
which implies, for every bath oscillator,
\begin{equation}
\label{eq:Cb-dot}
\dot{C}_b(t) =\frac{1}{2}\frac{\dot{\eta}(t)}{\eta(t)} C_b(t).
\end{equation}
The two kernels in (\ref{eq:gamma}) and (\ref{eq:chi}) then follow from 
\begin{equation}
 \begin{array}{lcll}
     \frac{C_{\omega}(t)C_{\omega}(s)f(\omega)}{m_{\omega}} &=& \frac{2\sqrt{\eta(t)\eta(s)}}{\pi}\omega^2 \qquad &\omega<\Omega  
 \end{array}    
\end{equation}
and are
\begin{eqnarray}
\label{eq:gamma-CL}
\gamma(t,s)&=& \frac{2\sqrt{\eta(t)\eta(s)}}{\pi}\int_0^{\Omega}\cos\omega(t-s)d\omega \\
\label{eq:chi-CL}
\chi(t,s)&=& \frac{\dot{\eta}(t)}{\pi}\sqrt{\frac{\eta(s)}{\eta(t)}} \int_0^{\Omega}\cos\omega(t-s)d\omega 
\end{eqnarray}
We now further assume that $C_b(t)$ change between zero and their full value on a time scale much longer than $\Omega^{-1}$
so that the integrals in (\ref{eq:gamma-CL}) and (\ref{eq:chi-CL})
can be approximated by delta functions. This leads to 
\begin{eqnarray}
\label{eq:gamma-CL-approx}
\gamma(t,s)&\approx& 2\eta(t)\delta(t-s) \\ 
\label{eq:chi-CL-approx}
\chi(t,s)&\approx& \dot{\eta}(t)\delta(t-s)
\end{eqnarray}
which means that the 
two terms inside the integral in (\ref{eq:FP-1}) evaluate to
\begin{eqnarray}
F_{fric} &=&- \eta(t) \dot{Q}(t) \label{eq:F-fric}   \\
F_{if}   &=& -\frac{1}{2}\dot{\eta}(t)Q(t) \label{eq:F-if}  
\end{eqnarray}
The first of the above is a standard friction force while the second is a time-dependent harmonic force, proportional to the time derivative of the 
friction coefficient.

Turning now to force term $\xi$ it varies rapidly in time and depends directly on the initial positions and momenta och the bath oscillators.
Its statistical properties follow from averaging over these initial positions and momenta, and  
it is immediate that $\hbox{E}\left[\xi(t)\right]=0$.
The second moment is
\begin{widetext}
\begin{eqnarray}
\hbox{E}\left[\xi(t)\xi(s)\right] &=& \int P_{GB}\left(\{q_b(t_i),p_b(t_i)\}\right) \prod_b d q_b(t_i)   d q_b(t_i) \left[\xi(t)\xi(s)\right] \nonumber \\
                                  &=& \sum_b C_b(t) C_b(s) \frac{1}{m_b\omega_b^2}\cos\omega_b(t-s) \approx 2 \eta(t) \delta(t-s)
\label{eq:xi-xi}
\end{eqnarray}
\end{widetext}
The random force $\xi$ 
therefore satisfies an Einstein relation with only the friction force $F_{fric}$. 

For the following discussion it is convenient to introduce, in analogy with the ``Sekimoto force''~\cite{Sekimoto98,sekimoto-book},
\begin{equation}
\label{eq:Fs}
F_S= \xi+F_{fric}+F_{if}
\end{equation}
It is also convenient to introduce an auxiliary quantity with dimension velocity
\begin{equation}
\label{eq:QB-dot}
\dot{Q}_B= -\frac{1}{2}\frac{\dot{\eta}(t)}{\eta(t)}Q
\end{equation}
such that 
\begin{equation}
\label{eq:Fs-2}
F_{fric}+F_{if}= -\eta\left( \dot{Q} - \dot{Q}_B\right).
\end{equation}
The interpretation of $\dot{Q}_B$ is that it is the time- and coordinate-dependent
velocity of the system such that the force $F_S$ from the bath on the system vanishes in expectation.
It is zero when the system-bath coupling is constant in time.

\section{Work and heat of stochastic thermodynamics at strong coupling}
\label{sec:StochThermo} 
We start by reviewing the derivation of Jarzynski's equality at strong coupling following~\cite{Jarzynski2004},
with adjustments arising from time-dependent system-bath coupling.
The system and the bath together consititute a closed system and the work 
done on the combined system is the change of the total energy 
\begin{equation}
\label{eq:H-tot}
\Delta H_{TOT} = \Delta H_S + \Delta H_B +\Delta  H_I + \Delta H_C.
\end{equation}
It is immediate that the combined system satisfies a Jarzynski equality 
\begin{equation}
\label{eq:JE}
\left< e^{-\beta \Delta H_{TOT}}\right>_{eq}  = e^{-\beta\Delta F_{TOT}}
\end{equation}
where the average is over the equilibrium state of the combined system
and $\Delta F_{TOT}$ is the change of total equilibrium free energy.
As we assume that the system and the bath are uncoupled at the
initial and the finite time $\Delta F_{TOT}$ is in fact here equal to
$\Delta F$, the free energy change of the system only.
For the same reason we can write
\begin{equation}
\label{eq:dW-1}
\delta W = \Delta H_{TOT} = \Delta H_S + \Delta H_B 
\end{equation}
the other two terms vanishing on the boundaries.
On the other hand we can also use Liouville's theory and write
\begin{equation}
\label{eq:dW-2}
\delta W = \int \partial_t H_S + \partial_t H_I  + \partial_t H_C  
\end{equation}
which has two extra terms compared to Jarzynski work when the system-bath coupling depends on time.
The bath Hamiltonian $H_B$ does not contribute to (\ref{eq:dW-2}) as it is independent of time.

The more recent development of heat in stochastic thermodynamics at strong coupling
starts from a Hamiltonian of the system at mean force~\cite{Seifert2016,Jarzynski2017}
\begin{equation}
\label{eq:H-meanforce}
{\cal H} = H_S + H_C -\beta^{-1}\log\left<e^{-\beta H_I}\right>_B   
\end{equation}
where the average is over the Boltzmann distribution of the bath only.
From this is derived an internal energy function at mean force defined as
\begin{equation}
\label{eq:E-meanforce}
E = {\cal H} +\beta\partial_{\beta}{\cal H}
\end{equation} 
For a bath of Hamiltonian oscillators linearly coupled to the system the average in ${\cal H}$ 
cancels with the counter-term $H_C$ so that we have simply
\begin{equation}
\label{eq:E-H-meanforce}
E={\cal H} = H_S.   
\end{equation} 
The change in this internal energy can be written as a time integral
\begin{equation}
\label{eq:E-1}
\Delta E=\int \partial_t H_S + \dot{Q}\partial_Q H_S +  \dot{P}\partial_P H_S.   
\end{equation} 
A definition of a path-wise heat functional which satisfies a First Law with
(\ref{eq:dW-1}) and  (\ref{eq:E-H-meanforce})
is then 
\begin{equation}
\label{eq:Q-meanforce}
\delta Q = \delta W - \Delta E = \Delta H_B.   
\end{equation}
In a round-about way we have thus arrived at the same notion of heat as in Introduction.
Using (\ref{eq:dW-2}) and (\ref{eq:E-1}) we can write $\delta Q$ as a time integral as
\begin{equation}
\label{eq:dQ-1}
\delta Q = \int \partial_t H_I  + \partial_t H_C  -  \dot{Q}\partial_Q H_S - \dot{P}\partial_P H_S      
\end{equation}

\section{Work and heat with a time-dependent Ohmic bath}
\label{sec:S}
We start by writing out  (\ref{eq:E-1}), (\ref{eq:dW-2}) and (\ref{eq:dQ-1}) explicitly for the model at hand:
\begin{eqnarray}
\label{eq:E-2}
\Delta E &=& \int \partial_t H_S + \frac{P}{M}\sum_b C_b\left(x_b - \frac{C_b Q}{m_b\omega_b^2}\right) dt \\
\label{eq:dW-3}
\delta W &=& \int \partial_t H_S - Q\sum_b \dot{C}_b \left(x_b - \frac{C_b Q}{m_b\omega_b^2} \right) dt \\
\label{eq:dQ-2}
\delta Q &=& \int \sum_b \left(-\dot{C_b}Q - \frac{P}{M}C_b\right)  \left(x_b - \frac{C_b Q}{m_b\omega_b^2}\right) dt 
\end{eqnarray}
For the change of internal energy it is seen that the sum over the bath oscillators
is the same as in (\ref{eq:P-dot}), and using (\ref{eq:Fs}) we can write
\begin{eqnarray}
\label{eq:E-3}
\Delta E &=& \int (\partial_t H_S) dt + F_S \circ dQ 
\end{eqnarray}
Equation (\ref{eq:E-3}) is the Jarzynski work ($\int \partial_t H_S dt$)
plus a term expressing the work done on the system from the total force $F_S$ that arises from system-bath coupling.
It can therefore read as the sum of the flow of energy into the system
from the external system and from the bath, which is the
same standard form in stochastic thermodynamics~\cite{Sekimoto98}.
Equation (\ref{eq:E-3}) only involves the time-dependence of the system-bath coupling through
the extra force $F_{if}$ entering $F_S$.

For the work functional, by the assumptions made in Section~\ref{sec:Ohmic}, we have 
\begin{eqnarray}
\label{eq:dW-4}
\delta W &=& \int (\partial_t H_S)dt  + F_S \circ dQ_B  
\end{eqnarray}
where $dQ_B=\dot{Q}_B dt$ and $\dot{Q}_B$ was introduced in (\ref{eq:QB-dot}).
The work is therefore the Jarzynski work plus the work done by the
Sekimoto force through the virtual increment of the system position $dQ_B$.  
The heat can similarly be written
\begin{eqnarray}
\label{eq:Q-1-def}
\delta Q &=& \int (-F_S) \circ \left(dQ - dQ_B \right) 
\end{eqnarray}
and is the work done by the Sekimoto reaction force ($-F_S$) through the 
difference between the actual increment $dQ$ and the virtual increment $dQ_B$.

An alternative way to divide up the different contributions to the heat which will be useful in 
Section~\ref{sec:Quantum} below is
\begin{eqnarray}
\label{eq:Q-2-def}
\delta Q &=& \int -\xi\circ (dQ-dQ_B) + \left(\frac{d}{dt}\left(\sqrt{\eta}Q\right)\right)^2 dt         
\end{eqnarray}
When $\eta$ is constant in time this reduces to Sekimoto's heat functional
$\int -\xi\circ dQ + \eta \dot{Q}^2dt$.

\section{The work done by the time-dependent system-bath coupling}
\label{sec:JE}
It has been seen above that both the heat and the work contain a term 
$\int F_S \circ dQ_B$
which is not reflected in the internal energy change at all.
Changing the system-bath coupling is a kind of external control 
acting on the system, which implies a kind of work. An original
feature 
is that this work is entirely dissipated
into heat (change of energy of the bath); nothing remains as change of internal energy.
Since $\dot{Q}_B$ has been assumed non-zero only near the initial and final
time this work is a kind of change of state function, and not a proper functional
depending on the whole path.

The first part of this work, corresponding to the first of the three terms in $F_S$, is
\begin{equation}
\label{eq:new-force-1}
\int F_{fric} \circ dQ_B = \int \frac{P}{2M}\, \dot{\eta}Q\, dt.
\end{equation}
Assuming that $\eta$ changes from zero to a finite value $\overline{\eta}$
over a short time period $\Delta t$ the force $F_{if}$ is going to dominate 
all the other forces in the Kramers-Langevin equation (\ref{eq:FP-1}).
This means that we have
\begin{equation}
P(t) = P_i - \frac{1}{2}\eta Q + {\cal O} (\sqrt{\Delta t})\qquad t\in [t_i,t_i+\Delta t ]
\end{equation}
while at the same time 
\begin{equation}
Q(t) = Q_i + {\cal O} (\Delta t)
\end{equation}
From this the contribution to (\ref{eq:new-force-1}), from the beginning
of the process, is $\frac{Q_iP_i}{2M}\overline{\eta} - \frac{Q_i^2}{8M}\overline{\eta}^2$,
and analogously just before the final time.
The contribution of this part is therefore the change of 
an auxiliary friction-dependent energy: 
\begin{eqnarray}
\label{eq:new-force-1-v3}
\int F_{fric} \circ dQ_B &=& \Delta V_{frict}+{\cal O} (\Delta t)\\
V_{frict} &=& -\frac{\overline{\eta}QP}{2M}+\frac{\overline{\eta}^2Q^2}{8M}
\end{eqnarray}
The second part of $\int F_S \circ dQ_B$ is
\begin{equation}
\label{eq:new-force-2}
\int F_{if} \circ dQ_B = \int \frac{1}{4}Q^2 \frac{\dot{\eta}^2}{\eta} dt 
\end{equation}
Similarly to above this can be written as 
a functional of $Q_i$, $P_i$ and the function $\eta$ in the interval $[t_i,t_i+\Delta t]$
and the same at the final time. 
We can therefore write
\begin{equation}
\label{eq:new-force-2}
\int F_{if} \circ dQ_B = A_i[Q_i,P_i,\eta] + A_f[Q_f,P_f,\eta] +{\cal O} (\Delta t)
\end{equation}
with two functionals $A_i$ and $A_f$.
The largest (most divergent) contributions to $A_i$ and $A_f$ are
$\frac{1}{4}Q^2\int \frac{\dot{\eta}^2}{\eta}dt$ which diverge as $(\Delta t)^{-1}$.
The third part of $\int F_S \circ dQ_B$ is
\begin{equation}
\label{eq:new-force-3}
\int \xi \circ dQ_B.
\end{equation}
This is a random variable which depends on the realization of $\xi$ just 
after the initial time and just before the final time of mean zero and  
variance 
\begin{equation}
\label{eq:new-force-3}
\left< \left(\int \xi \circ dQ_B\right)^2\right> = 2k_B T \left(A_i+A_f\right)
\end{equation}
The second and the third part of $\int F_S \circ dQ_B$ are hence potentially both
large, and when $\Delta t$ tends to zero they
can both be arbitrarily larger than standard Jarzynski work. 

Let us now consider the contribution of $\int F_S \circ dQ_B$
to Jarzynski equality, which we write
\begin{equation}
\label{eq:JE-fact}
\left< e^{-\beta \left(\int (\partial_t H_S)+F_S\circ dQ_B\right)}\right>_{fact. eq}  = e^{-\beta\Delta F}
\end{equation}
to emphasize that the initial equilibrium distribution is factorized between the system and the bath.
For given initial and final coordinates and momenta 
the first and second parts contribute simply $e^{-\beta \left(\Delta V_{fric}+ A_i+A_f\right)}$.  
The third part (\ref{eq:new-force-3}) on the other hand contributes
\begin{equation}
\label{eq:new-force-3}
\int \frac{1}{N} e^{-\frac{\beta}{2}\frac{x^2}{2\left(A_i+A_f\right)}}e^{-\beta x} dx = e^{\beta \left(A_i+A_f\right)}.  
\end{equation}
The (potentially divergent) contributions from the second and the third part 
hence cancel for each given initial and final coordinates and momenta (up to terms small as $\Delta t$),
and must therefore cancel in Jarzynski equality overall.

Combining this result and (\ref{eq:new-force-1-v3})
we have an alternative form of Jarzynski equality  
\begin{equation}
\label{eq:JE-fact-2}
\left< e^{-\beta \int (\partial_t H_S)-\beta \Delta V_{frict}}\right>_{fact. eq}  = e^{-\beta\Delta F}.
\end{equation}
This may be compared to the strong-coupling form of Jarzynki equality which holds
when $\eta$ is strictly constant in time:
\begin{equation}
\label{eq:JE-nonfact}
\left< e^{-\beta \int (\partial_t H_S)}\right>_{eq}  = e^{-\beta\Delta F_{TOT}}
\end{equation}
In above the average is over the system and the bath initially in joint equilibrium
with the terms $H_I$ and $H_C$ included, and the free energy change is
of the combined system and bath.
Pathwise the contributions to (\ref{eq:JE-fact-2}) and (\ref{eq:JE-nonfact})
from the Jarzynski work  $\int (\partial_t H_S)$ are the same up to terms ${\cal O}(\Delta t)$.
The differences between the two averages therefore stem from 
the additional term $\Delta V_{frict}$, and the different distributions over the initial conditions.

\section{Comparison to quantum entropy production}
\label{sec:Quantum}
In~\cite{AurellEichhorn} was computed the first-order change of the von Neumann entropy in a 
(quantum) bath of harmonic oscillators, between two measurements on the system
when the bath is initially in equilibrium at inverse temperature $\beta$.
The central quantity computed in that paper (Eq. 3 in~\cite{AurellEichhorn})
was
\begin{equation}
\label{eq:delta-S}
\overline{\delta S_q}=\frac{1}{P_{if}} \hbox{E}_{if}\left[\beta\hat{H}_B(t_f)\right] - \beta U(\beta) 
\end{equation}
where $i$ and $f$ denote the initial and final states of the system, $P_{if}$ is the transition probability
of  
and $\hbox{E}_{if}$
is the quantum expectation value of bath Hamiltonian at the
final time projected on these same states of the system, see (\ref{eq:a-Hf}) for the full formula in standard notation.
The bath is originally at equilibrium at inverse temperature $\beta$
and $U(\beta)$ is its equilibrium internal energy. 
The calculation of (\ref{eq:delta-S}) is a straight-forward generalization  
of the transition probability $P_{if}$ as computed by Feynman and Vernon
by integrating out the bath variables~\cite{FeynmanVernon}. 
In slightly
abbreviated form
where $\int_{if}$ stands for integration over the initial and final positions
of the system and projections on initial and final states, this quantity is
\begin{equation}
P_{if} = \int_{if} {\cal D}X {\cal D}Y e^{\frac{i}{\hbar}S_S[X]-\frac{i}{\hbar}S_S[Y]+\frac{i}{\hbar}S_i[X,Y]-\frac{1}{\hbar}S_r[X,Y]} \nonumber
\end{equation}
and $S_i$ and $S_r$ are real and imaginary parts of the Feynman-Vernon action.
The first main result of~\cite{AurellEichhorn} (Eq. 15) was that 
\begin{widetext}
\begin{equation}
\overline{\delta S_q}=\frac{1}{P_{if}} \left<{\cal I}^{(1)} + {\cal I}^{(2)}  + {\cal I}^{(3)}  \right>_{if} = \frac{1}{P_{if}}\int_{if} {\cal D}X {\cal D}Y e^{\frac{i}{\hbar}S_S[X]-\frac{i}{\hbar}S_S[Y]+\frac{i}{\hbar}S_i[X,Y]-\frac{1}{\hbar}S_r[X,Y]}
\left({\cal I}^{(1)}+ {\cal I}^{(2)}  + {\cal I}^{(3)}\right)
\label{eq:I1-I2-I3-average}
\end{equation}
\end{widetext}
where ${\cal I}^{(1)}$, ${\cal I}^{(2)}$ and ${\cal I}^{(3)}$ are three quadratic functionals. 
In this section will be discussed the classical limit of
$\left<{\cal I}^{(2)}  + {\cal I}^{(3)}  \right>_{if}$
and shown to be identical to the expected value of the classical heat discussed above in Section~\ref{sec:S}.

The change of expected bath energy can be written
\begin{equation}
\label{eq:delta-H-hat}
\hbox{E}_{if} \left[\Delta \hat{H}_B \right] = \hbox{E}_{if}\left[\hat{H}_B(t_f) \right] -\hbox{E}_{if}\left[\hat{H}_B(t_i)\right]
\end{equation}
where the last term is given in standard notation in (\ref{eq:a-Hi}).
The expression in (\ref{eq:delta-H-hat})
differs from (\ref{eq:delta-S}) by the absence of $P_{if}$
in the denominator and by the quantity subtracted at the initial time.
In fact, the expected energy of the bath energy measured at the 
initial time, conditioned on the future observation of the
final state $f$ of the system, is not the same as the
unconditioned equilibrium internal energy of the bath
times the transition probability,
a (simple) example of quantum retrodiction~\cite{PhysRevLett.111.160401,PhysRevA.91.062116}.
In appendix~\ref{a:conditioning}
the term $\left<{\cal I}^{(1)} \right>_{if}$,
also found in~\cite{AurellEichhorn},
is shown to be equal to the  
this discrepancy between 
$\hbox{E}_{if}\left[\hat{H}_B(t_i)\right]$
and $U(\beta)P_{if}$. 

Adapting to the setting of the present paper 
we from now on take the system-bath coupling coefficients $C_b$ to depend on time as in Section~\ref{sec:Zwanzig}
and we will divide the formula in~\cite{AurellEichhorn} by $\beta$ so that they express heat and not entropy production.
This leads to
\begin{eqnarray}
\label{eq:I2-definition}
{\cal I}^{(2)} &=& \int_{t_i}^{t_f} \int_{t_i}^{t_f} XY' h^{(2)}ds' ds \\
\label{eq:I3-definition}
{\cal I}^{(3)} &=&  \int_{t_i}^{t_f} \int_{t_i}^{t_f} XY' h^{(3)}ds' ds. 
\end{eqnarray}
where the corresponding two kernels are given in~\cite{AurellEichhorn} Eq.~16 (with one factor $\beta$ less):
\begin{eqnarray}
\label{eq:h2-definition}
h^{(2)}&=& i\sum_b\frac{C_b(s)C_b(s')}{2m_b}\coth(\frac{\beta\hbar\omega_b}{2})\sin\omega_b(s-s') \\
\label{eq:h3-definition}
h^{(3)}&=& \sum_b\frac{C_b(s)C_b(s')}{2m_b}\cos\omega_b(s-s')
\end{eqnarray}
In the Caldeira-Leggett limit these kernels tend to
\begin{eqnarray}
\label{eq:h2-CL}
h^{(2)}&\approx& -\beta^{-1}\frac{i}{\hbar} 2 \sqrt{\eta(s)\eta(s')} \frac{d}{d(s-s')}\delta(s-s')\\
\label{eq:h3-CL}
h^{(3)}&\approx& -\sqrt{\eta(s)\eta(s')} \frac{d^2}{d(s-s')^2}\delta(s-s')
\end{eqnarray}
Assuming that both $\eta$ and $\dot{\eta}$ vanishes at the initial and final time we can integrate 
by parts freely to find
\begin{widetext}
\begin{equation}
\label{eq:I2-I3-average-2}
\left<{\cal I}^{(2)}  + {\cal I}^{(3)}  \right>_{if} =
\left< \int \beta^{-1}\frac{i}{\hbar} \left( \left(\frac{d}{dt}(\sqrt{\eta}X)\right) (\sqrt{\eta}Y)-
                                            (\sqrt{\eta}X)\left(\frac{d}{dt}(\sqrt{\eta}Y)\right) \right) 
+\left(\frac{d}{dt}(\sqrt{\eta}X\right)\left(\frac{d}{dt}(\sqrt{\eta}Y\right) \right>_{if}
\end{equation}
\end{widetext}
In above the (quantum) expectation is the same as in (\ref{eq:I1-I2-I3-average})
and $X$ and $Y$ are the forward and backward quantum paths.
If we substitute for $X$ and $Y$ a classical path $Q$ and average over its probability distribution
the term $\left<{\cal I}^{(3)}  \right>_{if}$ in (\ref{eq:I2-I3-average-2}) is obviously the same
as the expectation value of the corresponding term in (\ref{eq:Q-2-def}).
The other term can be re-expressed as
\begin{widetext}
\begin{equation}
\label{eq:I2-average-3}
\left<{\cal I}^{(2)} \right>_{if} =
\beta^{-1}\frac{i}{\hbar} \left< \int
\eta \left(\frac{d}{dt}(X +Y) \right) \left(Y - X\right) +\frac{1}{2}\dot{\eta}\left(Y^2 - X^2\right)
+ \frac{1}{2}\frac{d}{dt}\left(\eta X^2\right) - \frac{1}{2}\frac{d}{dt}\left(\eta Y^2\right) \right>_{if}
\end{equation}
\end{widetext}
The separation in (\ref{eq:I2-average-3})
is analogous to the split into $\frac{2i}{\hbar}S_i^{mid}$ and $\Delta S_b$ in~\cite{AurellEichhorn}, Section 5,
with the difference that $\Delta S_b$ (the two last terms) now vanishes since $\eta$ is zero at the initial and final time.

To discuss the first two terms in (\ref{eq:I2-average-3}) we use the 
procedure of~\cite{AurellEichhorn}, Sections 6 \& 7.
The first step is to express the quantum averages through the Wigner transform
which for the first term in (\ref{eq:I2-average-3}) leads to the result in~\cite{AurellEichhorn}, Eq.~34.
For the second term in (\ref{eq:I2-average-3}) we have
that $\frac{1}{2}(Y^2-X^2)=Q\alpha$ where $Q=\frac{1}{2}(X+Y)$ is the average (eventually classical) path
and $\alpha=(Y-X)$ is the quantum deviation; a term $\frac{i\alpha}{\hbar}$ translates 
to a partial derivative with respect to momentum variable $P$ in the Wigner transform.
Collecting both terms we have hence 
\begin{widetext}
\begin{equation}
\label{eq:I2-average-3}
\left<{\cal I}^{(2)} \right>_{if} =
\beta^{-1}\int dsdQdP\,\, P(Q,P,Q_i,P_i)\left(-\frac{2\eta}{M}p\partial_p -\frac{\eta}{M} - \dot{\eta}q\partial_p\right) 
P(Q_f,P_f, Q,P)
\end{equation}
\end{widetext}
$P(q',p',q,p)$ in above is the Wigner function which in the Caldeira-Leggett limit tends to
the transition kernel of the classical stochastic process (\ref{eq:FP-1}). 

The second step is to compare the right-hand side of (\ref{eq:I2-average-3}) to the expectation value 
of the first term in (\ref{eq:Q-2-def})
$\left< \int -\xi\circ (dQ-dQ_B) \right>_{if}$. The expectation  
is conditioned
on the initial and final states which are assumed to be given by definite classical coordinates 
and momenta.
It is convenient to re-write this as
\begin{equation}
\label{eq:I2-average-4}
\int_{Q_i,P_i}^{Q_f,P_f}{\cal D}\hbox{Prob}(\hbox{path})\left[
\int \left(-\frac{P}{M}-\frac{\dot{\eta}}{2\eta}Q\right)\circ d\Xi\right]
\end{equation}
where $\hbox{Prob}(\hbox{path})$ is a weight over paths and
$d\Xi$ is the increment of the aggregated random force,
a Gaussian random variable with mean zero and variance $2k_B T \eta dt$.
One can therefore discretize the inner integral in (\ref{eq:I2-average-4}) and consider 
the contribution from the short time interval $[s,s+\Delta s]$ 
\begin{widetext}
\begin{equation}
\label{eq:I2-average-5}
\hbox{Term from Eq.~\protect\ref{eq:I2-average-4}}= 
\int dQdPdQ'dP' \left(-\frac{P'+P}{2M}-\frac{\dot{\eta}}{2\eta}Q\Delta s\right)\Delta \Xi\
P(\cdot,Q',P')P_{\Delta s}(Q',P',Q,P)P(Q,P,\cdot).
\end{equation}
\end{widetext}
The
short-time propagator of the Kramers-Langevin process is
\begin{widetext}
\begin{equation}
P_{\Delta s}(Q',P',Q,P) =\frac{1}{\cal N} e^{-\frac{1}{4k_B T\eta \Delta s}\left(P'-P+
\left(\partial_{Q}V + \eta\frac{p+p'}{2M}+\frac{1}{2}\dot{\eta}Q\right)\Delta s\right)^2}\delta\left(Q'-Q-\frac{P'+P}{2M}\Delta s\right)
\end{equation}
\end{widetext}
and has the property that
\begin{widetext}
\begin{equation}
\Delta\Xi P_{\Delta s}(Q',P',Q,P) = -2k_B T \eta \Delta s \partial_{P'} P_{\Delta s}(Q',P',Q,P) + {\cal O}\left(\Delta s^2\right)
\end{equation}
\end{widetext}
Integration by parts moves over the derivative with respect to $P'$ to respectively $P(Q_f,P_f,Q',P')$ and $\frac{P'+P}{2M}$
which gives the expression in (\ref{eq:I2-average-3}).
The same analysis through the Wigner function as outlined here can
of course also be applied to the last (simpler) term in
(\ref{eq:I2-I3-average-2}), with the same result as given above up to terms of order $\hbar^2$.

\section{Discussion}
\label{sec:Discussion}
The study of the interaction of a classical or quantum system with a bath
of harmonic oscillators has a long history, and 
it behoves the author to here motivate the need for another paper on the subject.

Consider the assumption of the Feynman-Vernon theory 
that the system and the bath are originally decoupled and the bath is at equilibrium.
If the coupling is not weak this is questionable
because if the system and the bath were in contact precisely at the initial
time they should have been so also slightly before, and then they could
not have been initially fully decoupled. 
Nevertheless, this assumption is needed to integrate out the bath variables
and arrive at the Feynman-Vernon open system development operator of the system 
only\footnote{Although the bath oscillators can be integrated out also
if the system and the bath are originally in equilibrium
together, the analysis is then considerably more complicated~\protect\cite{Grabert88}; 
that possibility has not been considered here.}.
It could therefore be argued that the Caldeira-Leggett theory 
of quantum Brownian motion~\cite{CaldeiraLeggett83a},
which is based on Feynman-Vernon, is only valid at weak coupling.
Indeed, it has been noted several times that the classical limit (Kramers-Langevin equation)
of an Ohmic bath has a delta-function force proportional to the friction
acting at the initial time, and that
the random and deterministic forces from the bath therefore do not obey
an Einstein relation. It has also been noted that it is a difference 
whether one assumes that the decoupled initial conditions pertain
exactly at the initial time or only slightly after.

To resolve these (and other) issues
Caldeira and co-workers in~\cite{daCosta2000}
investigated the possibility that
the bath is initially in equilibrium \textit{conditional}
on the position of the system \textit{i.e.}
with respect to Hamiltonian
$\sum_b \frac{p_b^2}{2m_b} + \frac{1}{2}m_b\omega_b^2 \left(q_b-\frac{C_b(t) Q}{m_b\omega_b^2}\right)^2$.
These initial conditions 
are mathematically possible and physically allowed, but raise the question of how
the system and the bath would find themselves in such a state.
If some control would have been exercised on the bath prior to the process
this control must have been aware of the position of the system, and
only a very exquisite procedure would have resulted in exactly the assumed initial state.

In this paper I have aimed to address these issees 
anew by allowing the system-bath coupling
depend on time. It is then not a problem to have the system and the bath  
initially decoupled, but the price to pay are new effects that arise
from the time-dependent friction. On the level of classical equations
these effects are just the extra force $F_{if}$ 
of (\ref{eq:F-if}) which when the switching
on and off of the bath-system coupling is fast reduces to the previously known delta-function force.
On the level of classical heat and work functionals
the situation is more involved and described
by (\ref{eq:dW-4}) and (\ref{eq:Q-1-def}).
There appears a part of the work denoted $\int F_S \circ dQ_B$
which is entirely dissipated into heat, which has support only 
at the beginning and the end of the process, and which is potentially quite large.
Nevertheless, it only makes a finite contribution to the
work functional that enters Jarzynski's equality, see (\ref{eq:JE-fact-2}).

On the quantum level of expected change of the bath energy the analysis can be carried out using
the Feynman-Vernon formalism, as done
previously for constant system-bath coupling in~\cite{AurellEichhorn}.
The classical limit of these expressions can then 
be seen to be the same as the classical heat functional with 
time-dependent friction. This is hence an example
of quantum-classical correspondence~\cite{PhysRevX.5.031038},
on the level of the expected heat.
The extra term $\Delta S_b$ found in~\cite{AurellEichhorn}
arises from using factorized initial conditions 
with a finite system-bath interaction present from the very beginning
of the process, and has therefore here been shown to be a kind of artefact.

\section*{Acknowledgments}
This paper was motivated by earlier joint work with
Prof.~Ralf Eichhorn whom I thank for many and stimulating discussions on the subject matter.
I thank Prof.~H.T. Quan and Dr.~Ken Funo for discussions and useful remarks, and the Institute of Theoretical
Physics at the Chinese Academy of Sciences (Beijing) for hospitality.
This research was supported by the Academy of Finland through its Center of Excellence COIN
and by the Chinese Academy of Sciences CAS
President’s International Fellowship Initiative (PIFI)
grant No. 2016VMA002.

\appendix

\section{An extra term found in~\protect\cite{AurellEichhorn}}
\label{a:conditioning}
We here discuss for completeness the first term in~\cite{AurellEichhorn} Eq.~15  
\begin{equation}
\begin{array}{ll}
{\cal I}^{(1)} &= \int^{t_f} \int^{s} (X-Y)(X'-Y')h^{(1)}ds' ds \\ 
\end{array}
\label{eq:I1-definition}
\end{equation}
with the kernel (a factor $\beta$ less compared to~\cite{AurellEichhorn})
\begin{equation}
\begin{array}{ll}
h^{(1)}&=-\sum_b\frac{C_b^2}{4m_b}\sinh^{-2}(\frac{\beta\hbar\omega_b}{2})\cos\omega_b(s-s')\\ 
\end{array}
\label{eq:h1-definition}
\end{equation}
As noted in~\cite{AurellEichhorn} (\ref{eq:I1-definition}) equals
$\partial_{\beta}\frac{1}{\hbar}S_r$ where $S_r$ is the real part of the 
Feynman-Vernon action. Writing the quantum expectation value of
${\cal I}^{(1)}$ in the Feynman-Vernon theory therefore means 
\begin{widetext}
\begin{eqnarray}
\left<{\cal I}^{(1)}\right>_{if} &=& \frac{1}{P_{if}}\int_{if} {\cal D}X {\cal D}Y e^{\frac{i}{\hbar}S_S[X]-\frac{i}{\hbar}S_S[Y]+\frac{i}{\hbar}S_i[X,Y]-\frac{1}{\hbar}S_r[X,Y]}\left(\partial_{\beta}\frac{1}{\hbar}S_r\right) = -\partial_{\beta}\log P_{if}
\label{eq:I1-average}
\end{eqnarray}
\end{widetext}
where the notation is as in Section~\ref{sec:Quantum} above.
The second equality follows because in the exponent only $S_r$ depends on $\beta$.

We want to relate the term in (\ref{eq:I1-average}) to the difference between
$\hbox{E}_{if}\left[\hat{H}_B(t_i)\right]$ and $P_{if}U(\beta)$,
where $U(\beta)$ is the unconditioned expected energy of the bath at inverse temperature $\beta$.
To do so we write more formally
\begin{widetext}
\begin{eqnarray}
\label{eq:a-Pif}
P_{if}                       &=& \left<f|\hbox{Tr}_B \left[ V \left(\rho_B^{eq}(\beta) \oplus |i\right>\left<i|\right) V^{\dagger} \right] |f \right> \\
\label{eq:a-Hi}
\hbox{E}\left[\hat{H}_B(t_i)\right] &=& \left<f|\hbox{Tr}_B \left[ V\left(\left(\hat{H}_B\oplus\mathbf{1}\right)\left(\rho_B^{eq}(\beta) \oplus |i\right>\left<i|\right)\right) V^{\dagger} \right] |f \right> \\
\label{eq:a-Hf}
\hbox{E}\left[\hat{H}_B(t_f)\right] &=& \left<f|\hbox{Tr}_B \left[ \left(\hat{H}_B\oplus\mathbf{1}\right) \left(V \left(\rho_B^{eq}(\beta) \oplus |i\right>\left<i|\right) V^{\dagger}\right) \right] |f \right> \\
\label{eq:a-U}
U(\beta)                    &=& \hbox{Tr}_B \left[ \hat{H}_B \rho_B^{eq}(\beta)  \right] 
\end{eqnarray}
\end{widetext}
where $V$ is the total unitary operation on the system and the bath, and $\rho_B^{eq}(\beta)$ is the initial (equilibrium) density matrix of the bath. 
Using that $\rho_B^{eq}(\beta)=e^{-\beta \hat{H}_B}/Z(\beta)$ and $-\partial_{\beta}\log Z = U$ we have 
$\hbox{E}_{if}\left[\hat{H}_B(t_i)\right]=-\partial_{\beta}P_{if} + UP_{if}$.
Therefore  
\begin{eqnarray}
-\partial_{\beta}\log P_{if} &=& \frac{1}{P_{if}}\hbox{E}_{if}\left[\hat{H}_B(t_i)\right] - U(\beta).  
\end{eqnarray}
which was to be shown.

\bibliography{fluctuations,AZZNotes}%
\end{document}